\documentclass{pasj00}
 \tenpoint

\SetRunningHead{J.H. Fan et al.}{Optical Variability for 4C 29.45}
\title{Optical Photometrical Observations and Variability for Quasar 4C 29.45}
\author{J. H. Fan$^{1,2}$, J. Tao$^{3}$, B. C. Qian$^{3}$, A. C. Gupta$^{4}$,
Y. Liu$^{1}$, \\
Y. H.  Yuan$^{1}$, J.H. Yang$^{5}$, H. G.
Wang$^{1}$, Y. Huang$^{1}$}
 \affil{
$^1$ Center for Astrophysics, Guangzhou University, Guangzhou 510400, China \\
$^2$ Physics Institute, Hunan Normal University, Changsha, China \\
$^3$ Shanghai Astronomical Observatory, Chinese Academy of Sciences, Shanghai 200030, China\\
$^4$ Tata Institute of Fundamental Research, Homi Bhabha Road, Colaba, Mumbai - 400 005, India \\
$^5$  Department of Physics and Electronics Science, Hunan
University of Arts and Science, \\
No. 170, West Dongting Avenue,
Changde 415000, P.R.China}

\Received{$\langle$reception date$\rangle$}
\Accepted{$\langle$acception date$\rangle$}
\Published{$\langle$publication date$\rangle$}
\begin{document}

\KeyWords{Galaxies: active - Galaxies: photometry -
Quasar:Individual: 4C 29.45 (1156+295)}

\maketitle

\begin{abstract}
We reported the result of long term optical variability of the
blazar 4C 29.45 (QSO 1156+295, Ton 599), carried out optical
photometric observations in Johnson V, Cousins RI passbands during
April 1997 to March 2002 using the 1.56 meter telescope of the
Shanghai Astronomical Observatory (SHAO) at Sheshan, China, compiled
the post-1974  optical photometric data of the blazar by combining
our new observations with the published optical data, and found
maximum variations in  different passbands: $\Delta$U = 4.41 mag,
$\Delta$B = 5.55 mag, $\Delta$V = 4.53 mag, $\Delta$R = 5.80 mag,
and $\Delta$I = 5.34 mag. The average  color indices are: U$-$B =
$-$0.54$\pm$0.18 mag, B$-$V = 0.56$\pm$0.21 mag, B$-$R =
0.93$\pm$0.18 mag, B$-$I = 1.51$\pm$0.24 mag, V$-$R = 0.44$\pm$0.15
mag and V$-$I = 1.03$\pm$0.23 mag. The post-1974 data give us an
excellent opportunity to search for the existence of possible
periodicity in the light curve. In search for periodicity in the R
passband light curve, we performed Jurkevich  test  and power
spectral (Fourier) analysis methods, and CLEANest algorithms to
remove false signals. We found possible periods of  3.55 and 1.58
years. The possible mechanism for the periodic variability was
discussed.
\end{abstract}

\section{Introduction}

BL Lac objects (BLs) and flat spectrum radio quasars (FSRQs) are collectively
known as blazars and belong to a subclass of radio-loud active galactic nuclei
(AGNs). FSRQs include the highly polarized quasars-HPQs, the optically
violently variables quasars-OVVs, core-dominated quasars-CDQ, and even
superluminal sources-SM.  Blazars are characterized by large and violent flux
variations, high and variable polarization, predominantly non-thermal emission
at almost entire electromagnetic spectrum and relativistic electrons tangled
with magnetic field in a relativistic jet nearly pointing towards our line of
sight (see \cite{urry95} for a review).  The variability time scales of
blazars range from a few minutes to several years and can be broadly divided
into three classes viz. micro-variability or intra-day variability (IDV), short
term variability and long term variability. Time scales for micro-variability,
short term variability and long term variability are a few minutes to less than
a day, a few days to less than a month and months to several years respectively
\citep{fan05}.

Photometric observations of blazars is an important tool to construct their
light curves and study the flux variation behavior on the diverse time scales.
Blazars are known to show occasional, unpredicted outbursts. The origin of
these outbursts is not yet well understood. It was found to be common that the
light curves of blazars show slow variations over the time scale of years
\citep{fan05}.

Searching for periodicity in the long term light curve of blazars is
an important tool to predict and catch major outbursts. After the
periodicity reported in the blazar 3C 120 \citep{jurk71}, some other
sources were also claimed to show periodicity in their light curves
(e.g. \cite{cipr03,cipr04,clem03,fan97,fan98a,fan02,fior04,kidg92,lain99,liu95,manc02,qian04,rait01,sill88,zhang99} and reference therein).
On the basis of binary black hole model given by \citet{sill88} for
the blazar OJ 287, they predicted that there should be an outburst
in the blazar in the year 1994. The predicted outburst was really
observed in OJ 287 in their OJ-94 project \citep{sill96}.

To establish the reality of periods, we need long duration data. The
lengthy data record required to demonstrate for periodicity depends
 on signal-to-noise ratio, systematic errors, regularity in times
of measurements and nature of the underlying variation. Perfectly
evenly sampled data on a perfectly periodic variation might not be
require for more than 1.5 periods to yield an accurate measurement
of the period \citep{fan98a}. But for most of the blazars, existing
data samples do not produce evenly sampled data due to irregular
telescope time allotment and sometimes cloudy sky condition in which
observing run could not be carried out. In this situation, data for
a much longer time span, say six times of the period, will be useful
to demonstrate the possible period. Periodicity is not simply a
random event but probably has some physical significance
(\cite{kidg92}, also see \cite{fan97}).

4C 29.45 is an OVV located at z = 0.729 \citep{wills83,glass83,wills92} and has
shown large variation \citep{branly96}, high and variable
polarizations ($P_{IR}=$28.06\%, $P_{opt}=$28\%, \cite{holmes84,mead90}) with
its position angle, fraction of the optical polarization varying very
dramatically \citep{hong99}. It has also been noticed that the source shows
short and long term optical flux variability \citep{noble96,reith01,ghosh00}.

This source is one of our prime candidates for the  long term
monitoring in the optical pass bands from SHAO. In this paper, we
presented the historical optical data of 4C 29.45 from the published
literature and from our new VRI passbands observations during the
period of 1997 to 2002. Section \ref{sec2} presents the observations
and data reduction technique, and results, in section \ref{sec3} we
report  the data from the published literature, in section
\ref{sec4} the periodicity analysis, in section \ref{sec5} the
optical and radio correlation, in section \ref{sec6} discussion and
conclusions are given.

\section{Observations and the data reduction}
\label{sec2}
\subsection{Observations}

Photometric observations of 4C 29.45 were carried out in BV Johnson and RI
Cousins pass bands using a liquid nitrogen cooled 200 series $1\mathrm{K}
\times1\mathrm{K}$ CCD detector at f/10 Cassegrain focus of the
$1.56\,\mathrm{m}$ telescope at SHAO, China. The each pixel of the CCD detector
projected on the sky corresponds to 0.25\arcsec\, in both the dimensions. The
entire CCD chip covers $\sim 4.3 \times 4.3\,\mathrm{arcmin}^{2}$ on the sky.
The seeing at SHAO usually varied from 1.2\arcsec\, to 1.5\arcsec\, during our
observing runs. Several bias frames were taken in each observing night, sky
flats were taken during twilight hours and some times dome flats were also
taken.

\subsection{Data reduction}

For each night, median bias and median flats were generated and used for image
processing (removing internal effect of the CCD detector). Image processing
(bias subtraction, flat field correction and cosmic rays removal) were done by
using the standard routines in the IRAF software. Processing of the data
(getting instrumental magnitude of standard stars and blazar in the field of
the blazar 4C 29.45) was done by multiple concentric circular aperture
photometric technique using APPHOT task of the IRAF software at Guangzhou
University, Guangzhou, China and at Shanghai Astronomical Observatory,
Shanghai, China using Pentium 4 computers.

We have used the method of our earlier papers
(\cite{fan01,qian04b}, see also \cite{gu06}) for determining the differential
magnitudes $\mathrm{O-S_1}$, $\mathrm{O-S_2}$ and $\mathrm{S_1-S_2}$
where O, $\mathrm{S_1}$ and $\mathrm{S_2}$ are blazar, standard star
1 and standard star 2 respectively in the blazar field.  We try to
choose $\mathrm{S_1}$ and $\mathrm{S_2}$ of nearly similar
brightness of O. In case, if we do not have $\mathrm{S_1}$ and
$\mathrm{S_2}$ of nearly similar brightness of O, we accept
$\mathrm{O-S_1}$, $\mathrm{O-S_2}$, and $\mathrm{S_1-S_2}\leq
3\,\mathrm{mag}$. This selection effect will not give large
difference in the uncertainty determination in the $\mathrm{O-S_1}$,
$\mathrm{O-S_2}$ and $\mathrm{S_1-S_2}$. When we have more than 2
standard stars in the blazar field satisfying our standard star
selection criterion, we calculate the differential magnitude between
these standard stars and the deviation in the differential
magnitude. Finally, we select only two standard stars which have the
least deviation. We investigate the variability of the blazar by
using the variability parameter C introduced by \citet{romero99}
(see also \cite{cellone00,fan01}). C is expressed as
\begin{equation}
    C={\frac{\sigma_{(\mathrm{O-S_{1}})}}{\sigma_{\mathrm{(S_{1}-S_{2})}}}}
\end{equation}
where $\sigma_{\mathrm{(O-S_{1})}}$ and $\sigma_{\mathrm{(S_{1}-S_{2})}}$ are
the scatters of $\mathrm{O-S_1}$ and $\mathrm{S_1-S_2}$ respectively. If
$\mathrm{C>3}$, then the target source is variable.

The rms errors are calculated from the two standard stars using the following
equation:
\begin{equation}
\sigma=\sqrt{{\frac{\sum(m_{i}-\overline{m})^{2}}{N-1}}}
\end{equation}
where $m_{i} = (m_{\mathrm{S_1}} - m_{\mathrm{S_2}})_{i}$ is the differential
magnitude of stars $\mathrm{S_1}$ and $\mathrm{S_2}$ while
$\overline{m}=\overline{m_{\mathrm{S_1}}-m_{\mathrm{S_2}}}$ is the differential
magnitude averaged over the entire data set, and $N$ is the number of the
observations in a particular night.

\begin{table*}
\caption{Standard stars name and magnitude in the field of the
blazar 4C 29.45. In the last column of the table, 1 and 2 refer
\citet{raiteri98} and \citet{smith95} respectively.}
\begin{center}
\begin{tabular}{lcccc}
\hline\hline
Star  Name    & V mag & R mag & I mag & Ref. \\
\hline
Star 1    & 13.39$\pm$0.05 & 13.01$\pm$0.02 &                 & 1 \\
Star 13   & 15.36$\pm$0.04 & 14.97$\pm$0.04 & 14.62$\pm$0.07  & 2 \\
Star 14   & 15.89$\pm$0.09 & 15.53$\pm$0.08 & 15.16$\pm$0.18  & 2 \\
Star 15   & 16.60$\pm$0.05 & 16.30$\pm$0.04 & 15.88$\pm$0.30  & 2 \\
\hline \hline
\end{tabular}
\end{center}
\label{tab1}
\end{table*}

Our calculations show that the $\mathrm{O-S_{1}}$, $\mathrm{O-S_{2}}$ and
$\mathrm{S_{1}-S_{2}}$ for all four standard stars in the blazar 4C 29.45 field
are not greater than 3 for V, R, and I band magnitudes.  Therefore, we could
easily select any two of these four stars (see Table \ref{tab1}) by using our
selection criterion for standard stars discussed above for data calibration
purpose. For data in V and R bands, we found star 1 and star 13 gave minimum
deviation, so, we used star 1 as $\mathrm{S_1}$ and star 13 as $\mathrm{S_2}$
for calibrating our V and R bands data.

Since star 1 does not have I magnitude, we determined the I magnitude of the
blazar in the following way. We calibrated the blazar separately by using star
13, star 14 and star 15. The calibrated I band magnitudes corresponding to star
13, star 14 and star 15 are called as m$_{13}^{I}$, m$_{14}^{I}$ and
m$_{15}^{I}$ respectively. Final I band magnitude of the blazar is reported by
using the value of ${\frac{1}{3}}(m_{13}^{I}+m_{14}^{I}+m_{15}^{I})$ and the
deviation of m$_{13}^{I}$, m$_{14}^{I}$ and m$_{15}^{I}$ as the uncertainty of
the magnitude. This method is also used to determine the magnitude and the
magnitude uncertainty of the blazar when there were $\leq$ 2 sets of the
observations present in a single night.

\begin{figure}
\begin{center}
    \FigureFile(70mm,50mm){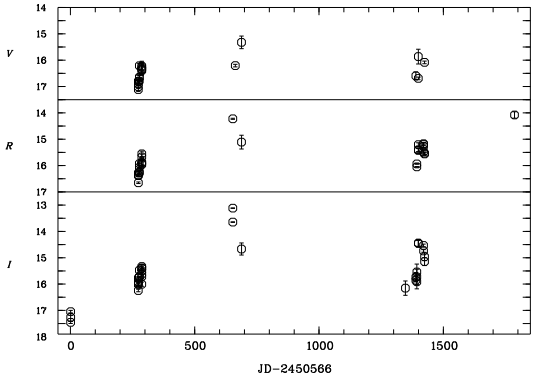}
\end{center}
\caption{Light curves of 4C 29.45. The upper panel is for V from
1998 to 2002, the middle panel is for R  from 1998 to 2002, and
the bottom panel is for I  from 1997 to 2002 respectively.}
\label{fan-pasj05-nl}
\end{figure}

\subsection{Observation results}

The results of our photometric observations of 4C 29.45 during January 1998 to
March 2002 in V and R bands, April 1997 to March 2002 in I band are given in
Table \ref{fan-pasj05-tabn} and plotted in Fig. \ref{fan-pasj05-nl}.  Column 1
in Table \ref{fan-pasj05-tabn} gives the Julian date, columns 2, 4, 6 represent
V, R, I magnitudes respectively, columns 3, 5, 7 give the uncertainty in V, R,
I data respectively.  The light curves clearly show the violent variability
behavior of the source. We found large variations (e.g. 2.08 mag (15.32 to
17.40 mag) in V, 2.57 mag (14.08 to 16.65 mag) in R and 3.13 mag (13.12 to
16.25 mag) in I band). From the I band observations, we found  the faintest
state of the source I = 17.45$\pm$0.05 on April 28, 1997.  Therefore, a
variation of 4.33 mag (13.12 to 17.45 mag) was found in I band during the
period of 1997 to 2002. Our observations presented in Table
\ref{fan-pasj05-tabn} indicate the source to be variable with the variability
parameter C being $C_{V}$ = 6.22, $C_{R}$ = 12.5, and $C_{I}$ = 5.92 for V, R
and I bands respectively. We found micro-variability in the source on Jan. 25,
1998 (JD 2450839), 4$\sigma$ variation of 0.3 mag in 1 hour in V band and
12$\sigma$ variation of 0.48 mag over 80 minutes in I band.

\begin{longtable}{lcccccc}
    \caption{Observational Journal for 4C 29.45}\label{fan-pasj05-tabn}
\hline\hline
 $JD date$ & V  & $\sigma_V$ & R  & $\sigma_R$& I  & $\sigma_I$ \\
 (1) &  (2)  &  (3)  &  (4)  & (5)  &  (6)  &  (7) \\
\hline
\endfirsthead
\hline
 $JD date$ & V  & $\sigma_V$ & R  & $\sigma_R$& I  & $\sigma_I$ \\
 (1) &  (2)  &  (3)  &  (4)  & (5)  &  (6)  &  (7) \\
\hline
\endhead
\hline
\endfoot
\hline\hline
\endlastfoot
2450567.118 &       &       &       &  &   17.45 & 0.04 \\
2450567.121 &       &       &       &       & 17.05   & 0.05 \\
2450567.15  &       &       &       &       &17.27   & 0.15 \\
2450839.297 &       &       &       &       &15.95 & 0.05 \\
2450839.309 &   16.80   &   0.07    &       & & &\\
2450839.316 &       &       &       &       & 15.77   & 0.05\\
2450839.326 & 16.90   &   0.07    & &       &       & \\
2450839.331 &       & &       & &   16.02   &   0.05 \\
2450839.335 &       &       & 16.28 &   0.03    &       & \\
2450839.339 &   17.12   &   0.07 & &       &       & \\
2450839.348 &       &       & &       & 15.99   &   0.05 \\
2450839.353 &       &       & 16.36   & 0.03 &       & \\
2450839.362 &       & &       &       & 15.91   & 0.05 \\
2450839.366 &       & &   16.38   &   0.03 &       & \\
2450839.37  &   17.01 &   0.07    &       &       & & \\
2450839.374 & &       &       &       &   16.25   &   0.05 \\
2450839.378 & &       &   16.65   &   0.03    &       & \\
2450839.382 & 16.82   &   0.07    &       &       &       & \\
2450843.263 &       &       &       &       &   15.72   & 0.01 \\
2450843.269 &       &       &   16.25   &   0.06    & & \\
2450843.28  &   16.21   &   0.09    &       &       & & \\
2450843.289 &       &       &       &       &   15.47   & 0.01 \\
2450843.293 &       &       &   15.93   &   0.06    & & \\
2450843.306 &   16.79   &   0.09    &       &       & & \\
2450843.314 &       &       &       &       &   15.73 &   0.01 \\
2450843.319 &       &       &   16.28   &   0.06 & &       \\
2450843.342 &       &       &   16.06   & 0.06    & &       \\
2450843.347 &   16.64   &   0.09    & &       & &       \\
2450853.196 &       &       &       & &   15.51   & 0.08    \\
2450853.201 &       &       &   15.67 &   0.1 & &       \\
2450853.212 &       &       &       & &   15.39   & 0.08    \\
2450853.229 &       &       &       & &   16.00   & 0.08    \\
2450853.239 &   16.22   &   0.18    & &       & &       \\
2450853.245 &       &       &       & &   15.62   & 0.08    \\
2450853.249 &       &       &   15.55 &   0.1 & &       \\
2450853.262 &   16.35   &   0.18    & &       & &       \\
2450853.272 &       &       &       & &   15.72   & 0.08    \\
2450853.3   &       &       &       & &   15.40   & 0.08    \\
2450853.305 &       &       &   15.96 &   0.1 & &       \\
2450853.311 &   16.40   &   0.18    & &       & &       \\
2450853.322 &       &       &   15.88 &   0.1 & &       \\
2450853.327 &   16.26   &   0.18    & &       & &       \\
2450853.332 &       &       &       & &   15.33   & 0.08    \\
2450853.336 &       &       &   15.91 &   0.1 & &       \\
2450853.341 &   16.33   &   0.18    & &       & &       \\
2451219.199 &       &       &       & &   13.12   & 0.01    \\
2451219.208 &       &       &       & &   13.65   & 0.01    \\
2451219.235 &       &       &   14.23 &   0.02    & &       \\
2451229.246 &   16.21   &   0.05 &       &       & &       \\
2451254.189 &       &       & &       &   14.67   & 0.23    \\
2451254.199 &       &       & 15.11   &   0.26    & &       \\
2451254.207 &   15.32   & 0.24    &       &       & &       \\
2451913.357 &       & &       &       &   16.15   & 0.27    \\
2451955.223 &       & &       &       &   15.74   & 0.18    \\
2451955.233 &       & &       &       &   15.86   & 0.05    \\
2451955.291 &   16.60 &   0.15    &       &       & &       \\
2451959.139 & &       &       &       &   15.91   & 0.27    \\
2451959.156 & &       &       &       &   15.72   & 0.29    \\
2451959.164 & &       &       &       &   15.54   & 0.30    \\
2451959.172 & &       &   16.05   &   0.01    & &       \\
2451959.187 & &       &   15.94   &   0.01    & &       \\
2451965.129 & &       &       &       &   14.45   & 0.15    \\
2451965.141 & &       &   15.21   &   0.07    & &       \\
2451965.168 & 15.86   &   0.28    &       &       & &       \\
2451965.187 &       &       &   15.43   &   0.07    & & \\
2451965.198 &       &       &       &       &   14.45   & 0.11 \\
2451965.203 &       &       &   15.40   &   0.07    & & \\
2451965.211 &   16.69   &   0.05    &       &       & & \\
2451986.152 &       &       &       &       &   14.53 &   0.05 \\
2451986.157 &       &       &       &       & 14.74   &   0.05 \\
2451986.206 &       &       &   15.27   & 0.05    &       & \\
2451986.217 &       &       &   15.48 &   0.07    &       & \\
2451986.23  &       &       & 15.20   &   0.05    &       & \\
2451986.254 &       & & 15.17   &   0.05    &       & \\
2451990.126 &       & & &       &   15.15   &   0.15    \\
2451990.127 &       & & 15.53   &   0.05    &       &       \\
2451990.14  &   16.08 & 0.05    &       &       &       & \\
2451990.153 & & &       &       &   14.96   &   0.15    \\
2451990.158 & & &   15.56   &   0.05    &       &       \\
2452352.148 & & &   14.08   &   0.12    &       &       \\
\end{longtable}

\section{Long-term optical photometry  data}
\label{sec3}

4C 29.45 is an important blazar and was often observed by different groups. We
compiled the optical data of the source from the available literature
\citep{ghosh00,glass83,kata00,mead90,rait99,sitko82,sitko91,smith87,smith88,vill97,webb88,wills83,xie94}
and our observations.  The data points from the literature are listed in Table
\ref{fan-pasj05-tabdata}, which shows clearly that most of observations are
from the papers by \citet{rait99}, \citet{vill97}, \citet{webb88}, and
\citet{wills83}.  The data in the papers by \citet{rait99} and \citet{vill97}
are given  in the standard B, V (Johnson), and R (Cousins) bands. In
\citet{webb88}, the U, B, and V data are in the standard Johnson UBV system,
they listed 145 B data for 4C 29.45.  For the data given in \citet{wills83}, we
compared some of them with those in \citet{webb88}. We chose the data with the
same observation time and found that those values given in the two literatures
are quite consistent and that there is no system difference. For data given in
\citet{glass83}, we compared those data that have the same observation time
with those in \citet{wills83} and found that those values given in the two
literatures are quite consistent. The data given in \citet{raiteri98} and
\citet{vill97} are in the same system and there is no system difference. We
also compared the data by \citet{kata00} with those by \citet{rait99}, and
found that there is no clear difference for those data with the similar
observation time.  The data given by our monitoring programme are also in
standard Johnson and Cousins system.  Therefore, the data compiled in this
paper are in the same passband system, there are no large system errors amongst
the data.  Some papers reported flux densities which were converted to
magnitudes by using the original magnitude-flux density formula. No
de-reddening correction was applied for the high galactic latitude (b$^{II} =$
80$^{\circ}$) of the object. We K-corrected all the magnitudes using the
formula
\begin{eqnarray}
    m = m^{ob} + 2.5 (1 - \alpha) \log (1 + z)
\end{eqnarray}
where $m$ and $m^{ob}$ are the magnitudes, in the source and the observer
frames, redshift z = 0.729 for 4C 29.45, and $\alpha$ the spectral index
($f_{\nu} \propto \nu^{-\alpha}$), here $\alpha=1.7$ was adopted from
\citet{fan96}.

\begin{table*}
\caption{Data Base (points) for 4C 29.45}
\begin{center}
\begin{tabular}{lccccccc}
\hline\hline
 Author  & Time(UT/JD) & U  & B & V & R & I & Note\\
 (1) &  (2)  &  (3)  &  (4)  &  (5)  &  (6)  &  (7) &(8) \\
\hline

\citet{glass83} & 1982 04 28 &     1  & 2      &  2     & 2 &2&flux
\\
 & 2445087.135 &       &       &       &  & & \\  \hline
\citet{mead90} & 1988 02 15--1988 02 17 & 3      & 3      & 3      &
2 & & flux \\
 & 2447206.348--2447208.348 &       &       &       &    & & \\
 \hline

\citet{kata00} & 1996 01 26--1997 02 01  &     &     &11&  & &mag\\
 &2450108.554--2450480.479 &       &       &       &    & & \\  \hline
 Present Work & 1998 01 25--2002 02 18  &     &     & &  32&  &mag
\\
 &2450839.302--2452352.148 &       &       &       &    & & \\  \hline
 \citet{rait99} & 1995 01 04--1996 04 13  &     &  56   &48 &
15&&mag\\
 &2449721.609--2450186.383 &       &       &       &    & & \\  \hline
 \citet{smith87} &  1983 01 08--1984 06 14  &  20    & 20      & 20& 20 & 18 &mag \\
 & 24453342.5 --2445865.5 &       &       &       &    & & \\  \hline
 \citet{vill97} & 1995 01 04--1995 05 09  &     &     &   & 31&
&mag\\
 &2449721.608--2449847.3594&       &       &       &    & & \\  \hline
 \citet{webb88} & 1980 05 09--1986 04 12  &     & 145   &   &  &
&mag \\
 &2444368.6--2446532.6&       &       &       &    & & \\  \hline
 \citet{wills83}& 1907 05 06--1983 07 15  &19     & 102   & 22 &16
&14&mag \\
 &2417702--2445531.178&       &       &       &    & & \\  \hline
 \citet{xie94} & 1990 04 20--1991 04 18  &      & 5    &4& 4 &4&mag\\
 & 2448001.075--2448364.318 &       &       &       &    & & \\  \hline

 \citet{raiteri98}& 1995 05 12--1995 12 07 &     &    &   &16  &  &fig \\
 &2451389.305--2452117.233&       &       &       &    & & \\
 \citet{raiteri98}& 1996 04 28--1999 06 02 &     &    &   &94  &  &
 \\
 &2450201.667--2451332.167&       &       &       &    & & \\ \hline
 \citet{reith01}& 1999 07 29--2001 07 26  &     &    &   &78  & &fig
\\
 &2451389.305--2452117.233&       &       &       &    & & \\ \hline
 \citet{webb88} & 1987 01 04--1990 01 28 &     &    &   &23  &  &fig \\
 &2446800--2447920&       &       &       &    & & \\ \hline\hline
\end{tabular}
\end{center}
\label{fan-pasj05-tabdata}
 \end{table*}

\begin{figure}
    \begin{center}
        \FigureFile(70mm,100mm){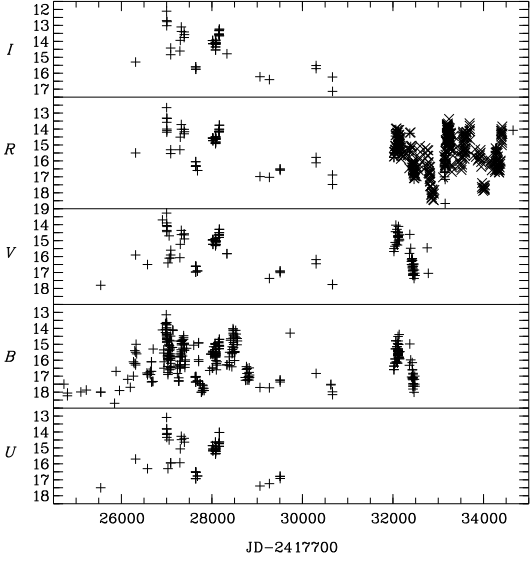}
    \end{center}
 \caption{Light curve of 4C 29.45. From the bottom to
the top, the panels show the UBVRI light curves respectively. The
crosses in R light curve represent the data from figures.}
\label{fan-pasj05-lc}
\end{figure}

The historical light curve covers a  time span of 95 years (1907-2002). The
data before 1974 are very sparse \citep{wills83}, the post 1974 early
photometric observations were made in UBVRI pass bands, but very recent
observations were mainly made in VRI pass bands. The historical light curves
were also shown by other authors \citep{rait99,reith01}.  Since the
data before 1974 are very sparse, in our following analysis, we only consider
the post-1974 UBVRI data as shown in  Fig.  \ref{fan-pasj05-lc}, which  covers
about 28 years (1974 to 2002).

\begin{figure}
    \begin{center}
        \FigureFile(70mm,50mm){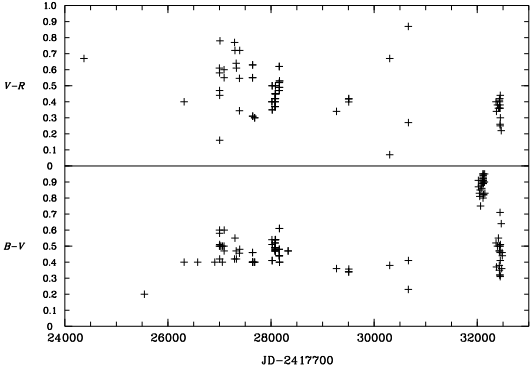}
    \end{center}
 \caption{Color variation. Upper panel for V-R, lower panel for B-V.}
\label{fan-pasj05-color}
\end{figure}

The long-term photometric data of 4C 29.45 show the maximum UBVRI variations in
the source-frame are as follows: $\Delta U = 4^m.41 \ (17.08-12.67)$, $\Delta B
= 5^m.55 \ (18.28-12.73)$, $\Delta V = 4^m.53 \ (17.38-12.85)$, $\Delta R =
5^m.80 \ (18.04-12.24)$, and $\Delta I = 5^m.34 \ (17.03-11.69)$. We also
determined the color indices from the compiled data and calculated the average
color index values. They are $U-B=-0.54\pm0.18$, $B-V= 0.56\pm0.21$, $B-R=
0.93\pm0.18$, $B-I= 1.51\pm0.24$, $V-R= 0.44\pm0.15$, $V-I= 1.03\pm0.23$ and
$R-I= 0.55\pm0.13$. The color index variation for V-R and B-V are presented in
Fig. \ref{fan-pasj05-color} for illustration.

The available data have also shown strong correlation between R and B (and V)
band data.
\begin{eqnarray}
R &=& (1.01\pm3.7\times10^{-4}) V - 0.65\pm8.34\times10^{-2}, \nonumber\\
R &=& (1.06\pm3.4\times10^{-4}) B - 1.92\pm9.0\times10^{-2}. \nonumber
\end{eqnarray}
with the correlation coefficients being 0.989.

\section{Periodicity analysis}
\label{sec4}

\begin{figure}
    \begin{center}
\FigureFile(70mm,50mm){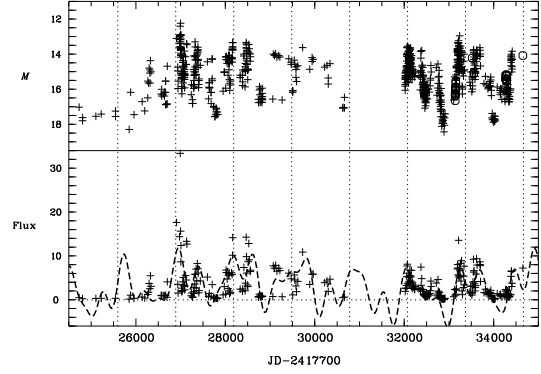}
\end{center}
\caption{R light curves. The upper panel shows the magnitude light
curve, the pluses indicate the data from  available literature while
the open circles our observations; the lower panel the 5-day
averaged R flux density (in units of mJy) light curve, the
number of 5-day avergaed data sample is 326. The dashed curve
is the theoretical result obtained by using the two strongest
CLEANest periods(Tab \ref{tab2}), the vertical dashed lines mean the
intervals at 3.55 years.} \label{fan-pasj05-fl}
\end{figure}

The light curves  gave us an opportunity to analyze  the periodicity
in the light curve. As we discussed, for the data compiled in this
paper there is no large system error, which will not affect the
long-term periodicity analysis result.  Since most data are
available in R band, we only used R fluxes for the periodicity
analysis.  Firstly, we converted magnitudes to fluxes by using the
magnitude-flux density formula. Secondly, in order to get nearly
evenly sampled data, we averaged the flux into 5 days intervals and
plotted the averaged light curve in the Fig. \ref{fan-pasj05-fl}.
The upper panel shows the R light curve (in magnitude), the lower
panel shows the 5-day averaged light curve (in flux in units of
mJy),  the number of 5-day averaged data sample is 326. The 5 days
interval is short enough in comparing to the long-term periods (in
years) and thus unlikely to distort much the long-term variation
behavior. In the following section, we will adopt Jurkevich method
and power spectral analysis to the averaged R data for periodicity
analysis.

\subsection{Jurkevich method}

\begin{figure}
    \begin{center}
\FigureFile(70mm,50mm){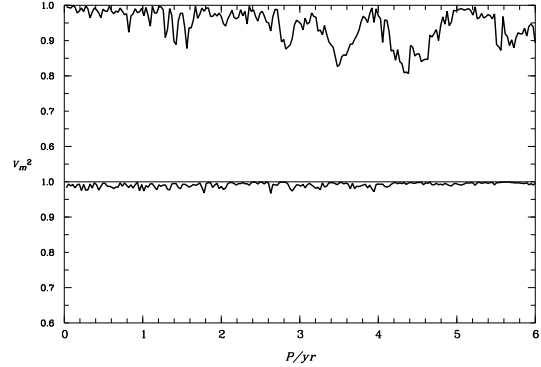}
\end{center}
\caption{Jurkevich analysis result. The upper panel is from the post-1974
light curve with $m=5$ while the lower panel is from the re-distribution of the
post-1974 light curve.} \label{fan-pasj05-jk}
\end{figure}

The Jurkevich method \citep{jurk71} is based on the expected mean square
deviation, it does not require an evenly distributed data sample.  So, it is
less inclined to generate spurious periodicity than the Fourier analysis. It
tests a run of trial periods around which the data are folded.  All data are
assigned to $m$ groups according to their phases around each trial period. The
variance $V_i^2$ for each group and the sum $V_{Sm}^{2}$ of all groups are
computed. For the whole data, one can calculate the variance $V_{h}^2$. By
virtue of the well known theorem on the addition variance, \citet{jurk71}
obtained that $V_{h}^2 > V_{Sm}^2$. We can then get the normalization
$V_{m}^{2}={\frac{V_{Sm}^2}{V_{h}^2}}$. If a trial period equals the true one,
then $V_m^2$ reaches its minimum. So, a ``good'' period will give a much
reduced variance relative to those given by other false trial periods and with
almost constant values.  The computation of the variances has been described in
the paper of \citet{jurk71}.

\citet{kidg92} introduced a fraction of the variance
\begin{eqnarray}
    f=\frac{1-V_m^2}{V_m^2}
\end{eqnarray}
where $V_m^2$ is the normalized value. In the normalized plot, a value of
$V_m^2 =1$ means f$=$0 and hence there is no periodicity at all. The best
periods can be identified from the plot: a value of $f \geq 0.5$ suggests  that
there is a very strong periodicity and a value of $f < 0.25$ suggests that the
periodicity, if genuine, is a weak one. A further test is the relationship
between the depth of the minimum and the noise in the ``flat'' section of the
$V_m^2$ curve close to the adopted period. If the absolute value of the
relative change of the minimum to the ``flat'' section is large enough as
compared with the standard error of this ``flat'' section, the periodicity in
the data can be considered as significant and the minimum as highly reliable
\citep{kidg92}.  Here we consider the half width at half minimum as the
``formal'' error as done by \citet{jurk71}.

It is possible that the distribution of the measurements, particularly for the
data concentrating in groups can give spurious periods, which are equal to the
interval of the time of the data groups. So, it is necessary to rule out the
spurious period detection. To do so, we used the Monte Carlo method to produce
synthetic ``measurements" and then adopted the Jurkevich method to the
synthetic ``measurements". If the trial period obtained from the true
measurements is also presented in the results obtained from the synthetic
``measurements", then the period is likely caused by the distribution and
should be discarded. By comparing the periods obtained from the real
measurements and those from the synthetic ``measurements", we can rule out the
spurious periods caused by the data distributions.  Therefore, the time for the
synthetic ``measurements" should be the same as that of the true observations.
The ``measurements" are produced as follows: the observing time is taken as the
time of each synthetic ``measurement" while the ``measurements" are then chosen
arbitrarily from the observed range of the actual observed data.  Namely, the
synthetic ``measurement" is simply a random re-distribution of the actual
observed data.

When the Jurkevich method was applied to the post-1974 averaged R fluxes ($m=5$
was adopted in our analysis), we found possible periods of $P_{1}$ =
0.82$\pm$0.03 (V$_{m}^{2}$ = 0.92, f = 0.08), $P_{2}$ = 1.43$\pm$0.15
(V$_{m}^{2}$ = 0.89, f = 0.12), $P_{3}$ = 2.82$\pm$0.07 (V$_{m}^{2}$ = 0.85, f
= 0.17), $P_{4}$ = 3.48$\pm$0.16 (V$_{m}^{2}$ = 0.81, f = 0.23), $P_{5}$ =
4.38$\pm$0.12 (V$_{m}^{2}$ = 0.80, f = 0.25) years.  Here the selection of P2
is from the three minima covering the interval  from 1.35 to 1.6 years  (see
Fig. \ref{fan-pasj05-jk}).

When we adopted the Jurkevich method to the random re-distribution of the
actual observed data for 4C 29.45, the result is shown in the lower panel in
Fig. \ref{fan-pasj05-jk}. By comparing  the analysis results from the true and
the synthetic ``measurements", we saw that all the periods appeared in the
upper panel in Fig.  \ref{fan-pasj05-jk} are not from the data distribution and
should be possible periods.

\subsection{Power spectral (Fourier) analysis}

We also performed a power spectral (Fourier) analysis, because it is
a powerful and common (well-studied) method to detect a periodic
signal, and gives some quantitative criterion for the detection of a
periodic signal.

In the case that the data are unevenly spaced in time, many attempts
of power spectral analysis have been made. In widespread use by
astronomers is the {\em modified periodogram} \citep{scargle82,horne86}, which is based on a least squares regression
onto the two trial functions, $\sin(\omega t)$ and $\cos(\omega t)$.
A superior technique is the {\em Date-Compensated Discrete Fourier
Transform}, or DCDFT \citep{ferraz81,foster95}, a least-squares
regression on $\sin(\omega t)$, $\cos(\omega t)$ and constant. The
DCDFT is a more powerful method than the {\em modified periodogram}
for unevenly spaced data, we adopted it to the R light curve, it can
be done as \citet{foster95} described.

The observed data $x(t_i)$ can define the data vector
\begin{equation}
    \left|x\right>=[x(t_1),x(t_2),\cdots,x(t_N)].
\end{equation}

First, defining the {\em inner product} of two vectors $f$ and $g$ as the
average value of the product $f^*g$ over the observation times $\{t_n\}$,
\begin{equation}
\left<f|g\right> = (\frac{1}{N})\sum^N_{n=1}f^*(t_n)g(t_n)
\end{equation}

A subspace are spanned by 3 trial functions $\phi_1(t)=1$
(constant), $\phi_2(t)=\sin(\omega t)$ and $\phi_3(t)=\cos(\omega
t)$. These 3 trial functions define a set of trial vectors,
\begin{equation}
    \left|\phi_\alpha\right>=[\phi_\alpha(t_1),\phi_\alpha(t_2),\cdots,\phi_\alpha(t_N)],~\alpha=1,2,3.
\end{equation}
The data vector $\left|x\right>$ can be projected onto the subspace
spanned by the $\left|\phi_\alpha\right>$ results in a model vector $\left|y\right>$ and a residual vector $\left|\Theta\right>$.
\begin{equation}
    \left|x\right>=\left|y\right>+\left|\Theta\right>
\end{equation}

The model vector $\left|y\right>$ is defined,
\begin{equation}
\left|y\right>=\sum_{\alpha}c_{\alpha}\left|\phi_{\alpha}\right>
\end{equation}

The $c_{\alpha}$ can be obtained, taking the inner product of each
trial vector $\phi_\alpha$ with the data vector $x$, we have
\begin{equation}
\left<\phi_\alpha|x\right>=\sum_{\beta} c_{\beta} \left<\phi_\alpha|\phi_\beta\right>=\sum_\beta S_{\alpha\beta}c_{\beta},
\end{equation}
which defines the $S$ matrix $S_{\alpha\beta}$. Inverting this matrix yields the coefficients,
\begin{equation}
c_\alpha=\sum_\beta S^{-1}_{\alpha\beta}\left<\phi_\beta|x\right>
\end{equation}
$s^2$ is the estimated data variance, it can be replaced by
$\sigma^2$. The power level of DCDFT is,
\begin{equation}
P_X(\omega) = \frac{1}{2}N[\left<y|y\right>-\left<1|y\right>^2]/s^2
\end{equation}

\begin{figure}
    \begin{center}
\FigureFile(70mm,50mm){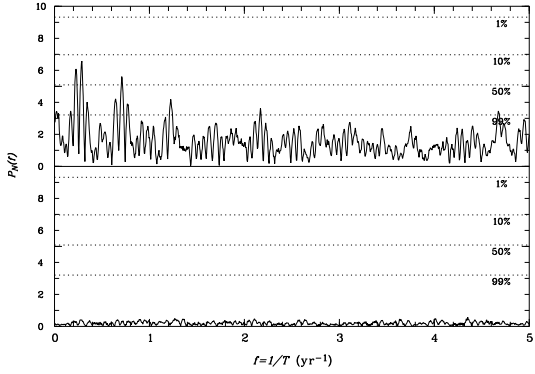}
\end{center}
\caption{The normalized Fourier analysis result for R light curve.
The upper panel is from the post-1974 light curve while the lower
panel is from the re-distribution of the same light curve. In the
upper panel, the highest peak is at $T=3.44\pm0.24$ years, with a
false probability of $FAP=0.136$, various false alarm probability
levels are marked. } \label{fan-pasj05-dcdft}
\end{figure}

We adopted the {\em False Alarm Probability}, $FAP$ \citep{horne86},
to give an auantitative criterion for the detection of a periodic
signal derived by DCDFT. \citet{horne86} introduced the False Alarm
Probability to deal with the modified periodogram.    In fact, the
False Alarm Probability (FAP) can deal with all kinds of Fourier
analysis method if the variations (mainly) consist of randomly
distributed noise. In Fig \ref{fan-pasj05-dcdft} and
\ref{fan-pasj05-clean} we noticed that the noise is almost randomly
distributed. It has been done by steps described here.

 First, the
power level of the periodogram is normalized by the total variance
$\sigma$,
\begin{equation}
P_{N}(\omega) = P_X(\omega)/\sigma^2
\end{equation}

The probability that $P_N(\omega_0)$ is of height $z$ or higher is
$Pr[P_{N}(\omega_0)>z]=e^{-z}$. Suppose that $z$ is the highest peak in a
periodogram that sampled $N_i$ independent frequencies. The probability that
each independent frequency is smaller than $z$ is $1-e^{-z}$, so the
probability that every frequency is lower than $z$ is $[1-e^{-z}]^{N_i}$. Thus,
the false alarm probability can be defined,
\begin{equation}
    FAP=1-[1-e^{-z}]^{N_i}
\end{equation}

For the independent frequencies, $N_i$, it is not too difficult to obtain $N_i$
by simple Monte Carlo, and the estimate of $N_i$ is not need be very accurate
\citep{press94}. We adopted $N_i=N_0/2.9$ in the case of our unevenly sampled
data, $N_0=326$ is the scale of our 5-day averaged data sample.

The DCDFT power spectral analysis are applied to the post-1974
averaged R data and the re-distributed observation data, the
resulting DCDFT is shown in Fig. \ref{fan-pasj05-dcdft}. The upper
panel is from the actual data while the lower panel is from the
re-distribution of the actual observed data. In the periodogram,
various false alarm probability levels are marked. The highest peak
is at $T=3.44\pm0.24$ years, with a false probability of
$FAP=0.136$. The second one is at $T=4.43\pm0.28$ years, with a
false probability of $FAP=0.233$, the third one is at
$T=1.42\pm0.05$ years, with a false probability of $FAP=0.326$. The
false probabilities of other peaks are all higher than 0.50.

\subsection{CLEANest analysis}

In the case of unevenly sampled time series analysis, irregular spacing
introduces myriad complications into the Fourier transform.  It can alter the
peak frequency (slightly) and amplitude (greatly), even introduce extremely
large false peaks.

\begin{figure}
    \begin{center}
\FigureFile(70mm,50mm){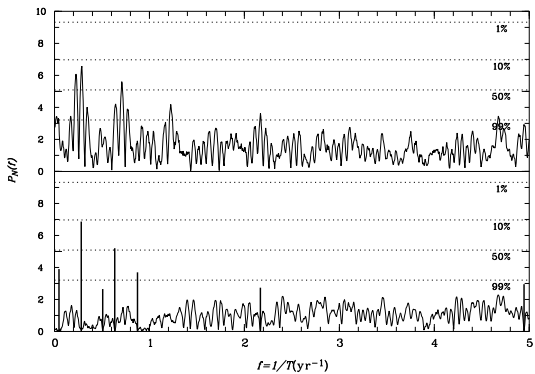}
\end{center}
\caption{The Fourier analysis and the CLEANest spectrum for R
light curve. The upper panel is the Fourier analysis from the
post-1974 light curve while the lower panel is the CLEANest spectrum
from the same light curve. In the lower panel, 7 CLEANest frequency
components (rough lines) and the residual spectrum are shown.
Various false alarm probability levels are marked. }
\label{fan-pasj05-clean}
\end{figure}

\citet{foster95} proposed the CLEANest analysis to clean false
periodicities, we also tried to do the CLEANest analysis. the
CLEANest algorithm can remove false peaks. First, the strongest
single peak and corresponding false components are subtracted first
from the original spectrum, then the residual spectrum is scanned to
determine whether the strongest remaining peak is statistically
significant. If so, then the original data are analyzed to fined the
pair of frequencies which best models the data, these 2 peaks and
corresponding false components are subtracted, and the residual
spectrum is scanned. The process continues, producing CLEANest
spectrum, until all statistically significant frequencies are
included. We assume that there are 7 independent frequency
components to clean observation data, the CLEANest spectrum is shown
in Fig \ref{fan-pasj05-clean}, and the results listed in Tab.
\ref{tab2}.

The variance of a frequency $Var(\omega)$ and the variance of the amplitude of
the given frequency $Var(P)$ can be estimated by \citet{foster96}
\begin{eqnarray}
    Var(\omega)&=&\frac{24\sigma_{res}^2}{NA^2T^2}\\
    Var(P)&=&\frac{2\sigma_{res}^2}{N}
\end{eqnarray}
where $\sigma_{res}$ is the variance of the residual data, $A$ is
the amplitude of the given frequency and $T$ is the total time span.
The $\sigma_{res}^2$ is estimated by
\begin{equation}
    \sigma_{res}^2=\frac{NV_{res}}{N-3f-1},
\end{equation}
where $V_{res}$ is the variance of residual data,
$V_{res}=\left<res|res\right>-\left<1|res\right>$, and $f$ is the
number of discrete frequencies.

We also introduce the False Alarm Probability to deal with CLEANest
frequency components, because of the same definition of amplitude.
$FAP$ are also listed in Tab \ref{tab2},  the strongest component at
$T=3.55\pm0.02$ years, with a amplitude $6.87\pm0.63$ and with a
false probability of $FAP=0.11$, the second one at $T=1.58\pm0.01$
years, with a amplitude $5.21\pm0.63$ and with a false probability
of $FAP=0.46$. The false probabilities of other components and
residual spectrum are more higher than 0.50. Notably, FAP means a
strong enough signal with a small probability to be false, for a
weak signal with stronger signal, it is can be true but with a large
FAP.

\begin{table*}[ht]
    \caption{Seven CLEANest frequency components for blazar 4C 29.45.}
    \begin{center}
    \begin{tabular}{cccc}
        \hline\hline
        Frequency & Period &Coeff. & FAP \\
    (cycles/yr) & (yr) & & \\\hline
        0.28 & $3.55\pm0.02$ & $6.87\pm0.63$ & 0.11\\
        0.63 & $1.58\pm0.01$ & $5.21\pm0.63$ & 0.46\\
        0.05 & $21.1\pm0.51$ & $3.90\pm0.63$ & 0.90\\
        0.88 & $1.14\pm0.01$ & $3.70\pm0.63$ & 0.94\\
        5.00 & $0.20\pm0.01$ & $2.96\pm0.63$ & 1.00\\
    2.17 & $0.46\pm0.01$ & $2.74\pm0.63$ & 1.00\\
        0.51 & $1.97\pm0.01$ & $2.66\pm0.63$ & 1.00\\\hline\hline
    \end{tabular}
    \end{center}
     \label{tab2}
\end{table*}

\subsection{Periodicity analysis results}

 We listed periodicity analysis results derived by JV method, DCDFT
and CLEANest in Tab \ref{tab_2}.
  Jurkevich method gives 5
possible periods, a closer look at all these 5 possible periods
showed that $P_{3} \sim 2 P_{2}$.  Therefore, $P_{3}$ are harmonics
of the $P_{2}$, so $P_{2}$ is a possible period along with $P_{1}$,
$P_4$ and $P_{5}$.  From power spectral (Fourier) analysis, we got 3
possible periods at 3.44 years($\sim P_4$), 4.43 years($\sim P_5$)
and 1.42 years($\sim P_2$), with false alarm probabilities lower
than 0.40. Using two different analysis methods, we got the same
results. These signals may be influenced by irregular spacing, the
peak frequencies and amplitudes are altered some. From CLEANest
analysis, we correct the influence and derived 2 periods at 3.55
($\sim P_4$) years and 1.58 ($\sim P_2$) years, with FAPs lower than
0.50 and amplitudes more than three times variance. So, we could
conclude that there are two possible periods of 1.58 and 3.55 years.
 We marked the light curve using the interval of 3.55 years, and
showed the theoretical light curve obtained by using the two
strongest CLEANest periods (3.55 and 1.58 years) (see Tab
\ref{tab2}) in Fig \ref{fan-pasj05-fl}.

\begin{table*} \caption{Periodicity analysis results.} \label{tab_2}
\begin{center}
\begin{tabular}{lccccc}
\hline\hline
& $P_1$ & $P_2$ & $P_3$ & $P_4$ & $P_5$ \\
&  (yr) &  (yr) &  (yr) & (yr)  & (yr)  \\\hline
JV & $0.82\pm0.03$ & $1.43\pm0.15$ & $2.82\pm0.07$ & $3.48\pm0.16$ & $4.38\pm0.12$ \\
DCDFT & & $1.42\pm0.05$ & & $3.44\pm0.24$ & $4.43\pm0.28$ \\
CLEANest & & $1.58\pm0.01$ & & $3.55\pm0.02$ & \\\hline\hline
\end{tabular}
\end{center}
\end{table*}

\section{Optical and Radio Correlation}
\label{sec5}

\begin{figure}
    \begin{center}
\FigureFile(70mm,80mm){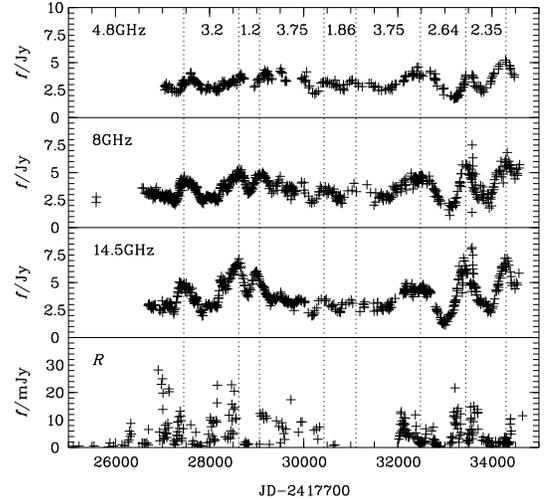}
\end{center}
\caption{Radio and optical light curves.  The panels from the top
to the bottom are the 4.8GHZ radio light curve, 8GHz light curve,
14.5GHz light curve and optical R light curve.}
\label{fan-pasj05-rop}
\end{figure}

4C 29.45 is also variable in the radio bands, light curves covering the period
from 1980 to 2002 were shown by \citet{hong04}. It shows clear  outbursts
at the radio and optical bands as shown in Fig. \ref{fan-pasj05-rop}. Is there
any correlation between the radio and  optical variation for the source?  To
study this, we adopted the discrete correlation function (DCF) method to the
radio and the optical data. The DCF method, which was described in detail by
\citet{edel88} \citep{fan98b}, is intended for analysis
of the correlation of two data sets. This method can indicate the correlation
of two variable temporal series with a time lag.  It can be done as follows.

Firstly, we have calculated the set of unbinned correlation (UDCF) between data
points in the radio and optical data streams a (for the optical data) and b
(for the radio data), i.e.
\begin{eqnarray}
 {UDCF_{ij}}={\frac{ (a_{i}- \bar{a}) \times (b_{j}- \bar{b})}{\sqrt{\sigma_{a}^2 \times
 \sigma_{b}^2}}},
\label{UDCF}
\end{eqnarray}
where $a_{i}$ and $ b_{j}$ are points in the data sets, $\bar{a}$ and $\bar{b}$
are the average values of the data sets, and $\sigma_{a}$ and $\sigma_{b}$ are
the corresponding standard deviations. Secondly, we have averaged the points
sharing the same time lag by binning the $UDCF_{ij}$ in the suitable sized
time-bins in order to get the $DCF$ for each time lag $\tau$:
\begin{eqnarray}
    {DCF(\tau)}=\frac{1}{M}\Sigma \;UDCF_{ij}(\tau), \label{DCF}
\end{eqnarray}
where $M$ is the total number of pairs. The standard error for
each bin is
\begin{eqnarray}
    \sigma (\tau) =\frac{1}{M-1} \{ \Sigma\; [
    UDCF_{ij}-DCF(\tau) ]^{2} \}^{0.5}.  \label{sigma}
\end{eqnarray}

\begin{figure} \begin{center} \FigureFile(70mm,90mm){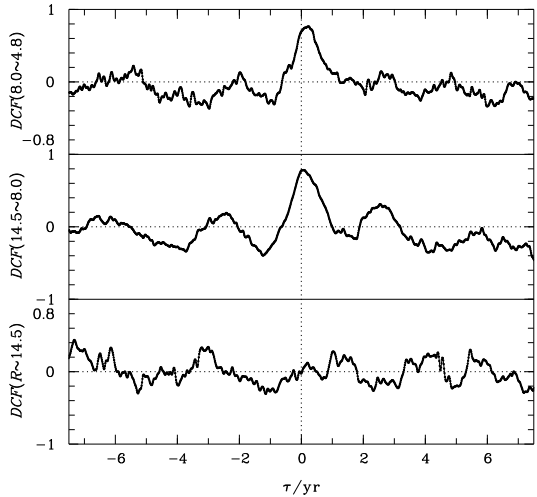}
\end{center}
\caption{DCF result for the post-1974 data. The top panel is for
the 8.0 GHz and the 4.8GHz radio data, the
middle one for the 14.5GHz and 8GHz data, the bottom panel for
the DCF result between optical R data and the 14.5GHz radio data(DCF bin: 30 days).}
\label{fan-pasj05-ro-dcf}
\end{figure}

When Eqs. (\ref{UDCF}) to (\ref{sigma}) are applied to the post-1974 optical R
and the radio data as shown in  Fig. \ref{fan-pasj05-rop}, the resulting DCF
result is obtained and shown in Fig.  \ref{fan-pasj05-ro-dcf}.  For the optical
and the radio variation, there is a very weak correlation with the optical
variation leading the 14.5GHz radio variation by 520--560 days.   For
comparison, the DCF results for radio bands are also calculated and shown in
Fig.  \ref{fan-pasj05-ro-dcf}.  For the radio band light curves, there is clear
correlation between the 14.5 GHz light curve and the 8.0GHz as well as the
8.0GHz light curves with the 4.8GHz variation leading the 4.8GHz by about 0
$\sim$ 280 days.

\section{Discussion and conclusion}
\label{sec6}

The compiled data of the OVV 4C 29.45  indicate  large optical flux variations.
From our observations, we found a faint state of the source I = 17.45$\pm$0.05
on April 28, 1997. \citet{xie94} also reported a faint state of the source I =
17.14 in their observations in 1991. Our data show that the source was even 0.3
mag fainter in I in 1997 compared to 1991.  In addition, \citet{vill99}
monitored the source during the period of January 04, 1995 to May 27, 1998 and
found a variation of 5.1 mag. The faintest state of the source was detected by
them on May 21, 1997 (R = 18.46). Present observations and observations by
\citet{vill99} show that the source was in a low-state in the first half of the
year 1997.

For periodicity analysis, according to a paper by \citet{kidg92}, we
found that all the periods determined from Jurkevich method are
weak. Further consideration indicated that the physically possible
periodicity is $P_{1}$, $P_{2}$, $P_4$ and P$_{5}$.  When the power
spectral (Fourier) analysis is adopted to the R flux, the analysis
result clearly shows that there are three possible periods at 3.44
years, 4.43 years and 1.42 years, with false alarm probabilities
  being smaller than 0.40. When we used the  CLEANest analysis to
correct the influence of irregular spacing, we obtained two
corresponding periods at 3.55 years and 1.58 years.

For comparison, we also used the DCF method for periodicity analysis as we did
in our previous paper \citep{fan02}, and found period signs at 3.2$\pm$0.08
and 4.15$\pm$0.15 years. However, there is no sign of 1.4 or 1.6 year period in
the analysis result obtained by DCF method. For the DCF method itself, we found
that it is a powerful periodicity analysis method if there is only one period
in the light curve. However, if  more than one period is present, DCF analysis
will dilute the sign of other periods. Therefore, DCF is an excellent method
for single period analysis but for multiple periods. We think that the DCF
analysis result should be compared with other methods.

It is interesting that its infrared light curves also show an interval of 3.2
years in our previous work \citep{fan99a},
  which is marginally consistent with
the $3.55\pm0.1$ years period. So, we can say that there are two possible
periods of 1.58 and 3.55 years in the light curve.

There are some popular models which can explain the long-term periodic
variations in AGNs viz. Double black hole model, the thermal instability model,
and the perturbation model
\citep{sill88,meyer84,abra99,romero03,rieger04,xie04,wu05}.

The thermal instability of slim accretion disk is a suitable model to see the
long term variability behavior. This model can explain ultra-short period
variable with a shorter duration of quiescence phase. Periodicity time scale of
a few years can not be produced by the thermal instabilities
\citep{fan99b}. Variability behavior of 4C 29.45 reported here has shown
the time scale of a few years which can not be supported by the model. So, this
model is ruled out for the explanation for the present work.

The perturbation model based on the accretion disk pulsations, the presence of
a single dominant hot spot on the disk. The model can show the periodicity in
blazar on timescales of a week or less. Here we reported the periodicity of the
time scale of a few years which can not be supported by this model.

\citet{sill88} suggested a double black hole model for the blazar
and predicted that the outburst in the blazar OJ 287 should occur in
1994. The predicted outburst was really observed in their OJ-94
project \citet{sill96}. The model given by \citet{sill88} is a well
established model for the long term periodic variation and the
prediction of next outburst in blazars. The periodicity reported
here can be successfully explained by the model.

4C 29.45 shows variation in the radio as well as in the optical band, the
variations in the two bands show weak correlation with the optical variation
leading the 8GHz radio one by 520-560 days. The variations at the three
frequencies (4.8GHz, 8GHz, and 14.5 GHz) are correlated with the variations at
higher frequencies leading the ones at lower frequencies.

In the present paper, we reported our optical (VRI) monitoring
results for the blazar 4C 29.45 during the period of 1997 to 2002.
The historic data were also compiled from the available literature.
The data presented in the paper showed that the source shows large
variations in UBVRI  bands. In addition, we adopted the periodicity
analysis methods to the R band light curve and found that there are
possible variation  periods of 1.58 and 3.55 years.


We thank the referee for the constructive comments and suggestions, and Dr. V.
R. Chitnis for correcting English of the paper. This work is partially
supported by the National 973 project (NKBRSF G19990754), the National Science
Fund for Distinguished Young Scholars (10125313), the National Natural Science
Foundation of China (10573005), and the Fund for Top Scholars of Guangdong
Provience (Q02114). We also thank the financial support from the Guangzhou
Education Bureau and Guangzhou Science and Technology Bureau. ACG's efforts are
partially supported by the Department of Atomic Energy, Govt. of India.


\end{document}